\documentclass[twocolumn,showpacs,preprintnumbers,amsmath,amssymb]{revtex4}
\usepackage{epsfig}
\usepackage{graphicx}
\usepackage{isolatin1}

\renewcommand{\L}{\mathcal{L}}               
\newcommand{\eins}{1\!\! 1}                  
\newcommand{\proj}{\mathbb{P}}               
\renewcommand{\i}{\mathrm{i}}                
\newcommand{\e}{\mathrm{e}}                  
\newcommand{\ket}[1]{|#1\rangle}             
\newcommand{\bra}[1]{\langle #1|}            
\renewcommand{\d}[1]{\mathrm{d}#1}           
\renewcommand{\cal}[1]{\mathcal{#1}}         
\newcommand{\mt}[1]{\mathrm{#1}}             

\begin{document}

\title{Blinking molecules:  Determination of Photo-Physical Parameters
from the \\ Intensity Correlation Function}   
\author{Gerhard C. Hegerfeldt}
\email{hegerf@theorie.physik.uni-goettingen.de}
\author{Dirk Seidel}
\affiliation{Institut für Theoretische Physik, Universität Göttingen,
  Bunsenstrasse 9, 37073 Göttingen, Germany}

\begin{abstract}
An explicit expression is given for the correlation function of blinking
systems, i.e.~systems exhibiting light and dark periods in their fluorescence. 
We show through the example of terrylene in a crystalline
host that it is possible to determine by means of this explicit
expression photo-physical parameters, like
Einstein coefficients and the mean light and
dark periods by a simple fit.  In addition we obtain further
parameters like the frequency of the various intensity periods and the
probability density of photons scattered off the host crystal. It
turns out that this approach is simpler and allows 
greater accuracy than previous procedures. 
\end{abstract}

\maketitle

\section{Introduction}

Since about twenty years it has been possible to observe and study the
fluorescence of single ions in Paul traps \cite{Neuhauser-PRA-1980}
and, more recently, also 
of single molecules embedded in a crystal
\cite{Moerner-PRL-1989,Orrit-PRL-1990, Tamarat-JPC-2000}. In the
presence of 
metastable triplet levels the molecule can exhibit light and dark
periods in its fluorescence (``blinking'') \cite{Basche-Nature-1995,
  Vogel-JPC-1995}, just as ions in the $V$ or $\Lambda$ 
configuration \cite{Dehmelt-BAP-1975,Nagourney-PRL-1986,
  Bergquist-PRL-1986, Sauter-PRL-1986, Sauter-OC-1986, 
  Diedrich-PRL-1987, Itano-PRL-1987, Itano-PRA-1988}. 
Spectroscopy of single molecules is of fundamental importance both for basic quantum
mechanical aspects like antibunching, quantum jumps or the dynamical
Stark effect \cite{Basche-PRL-1992,Basche-Nature-1995,
  Tamarat-PRL-1995}, as well as for applications in chemical and
biophysics, cf. e.g. \cite{Tamarat-JPC-2000, Schwille-CBB-2001}. For these applications a precise
knowledge is required of the parameters characteristic of
fluorescence, like Einstein coefficients and the mean duration of intensity
periods. Often these parameters cannot be calculated directly but
have to be determined indirectly from suitable experimentally
accessible quantities.

One of the most important statistical quantities for the description
of fluorescing quantum systems is the intensity correlation function,
$g(\tau)$ \cite{Mandel-book}. Its behavior for small times $\tau$ yields insight into the
nature, classical or quantum, of the photon statistics i.e. bunching
$(g(0) > 1)$ or antibunching $(g(0) < 1)$, respectively, and it shows
the effect of Rabi oscillations. In addition it may yield information
on light and dark periods through long-time correlations. The
importance of $g(\tau)$ originates from its easy experimental
accessibility as it does not depend on the detector efficiency.
To obtain a deeper physical understanding of this important quantity,
analytical results are often useful. By means of our recently proposed
approach for an analytical calculation of photon correlation functions
of arbitrary blinking quantum systems \cite{Hegerfeldt-QSO-2002} we
derive in this paper, for the first time, an explicit expression for
the correlation function of the 
four-level scheme used to describe the fluorescence of single molecular
systems like terrylene or pentacene. In contrast to previous
work, our result depends directly on the relevant photo-physical
parameters, such as Einstein coefficients, Rabi frequencies or the
mean duration of the light and dark periods, with no further
unknown quantities.
This allows an easy determination of these in
general unknown molecular parameters by a single fit to experimental
data, where in comparison to previous approaches the fit is greatly
facilitated through the explicit knowledge of $g(\tau)$, as will be
shown in the present paper for the example of a terrylene molecule embedded in
p-terphenyl. This intensity correlation function has recently been
measured in Refs.~\cite{Kummer-CPL-1994,
  Kummer-JPC-1995, Basche-BBG-1996}. In this paper we first derive an explicit
expression of $g(\tau)$ for terrylene, taking into account both its
metastable levels, and then fit $g(\tau)$ to the data of
Ref.~\cite{Basche-BBG-1996}. From this we obtain the electronic
Einstein coefficients, 
the mean light and dark periods and further fluorescence parameters
and compare our results with those in the literature.

\section{The intensity correlation function}

The energy-level configuration of planar hydrocarbons, such as
terrylene, can be described by an electronic three-level system \cite{Birks-book, Bernard-JCP-1993},
with a singlet ground state, $\ket{1}$, an excited singlet state, $\ket{3}$, and
a triplet state, $\ket{2}$. Due to the interaction between the magnetic
moments of the electron spins the metastable triplet state of terrylene
splits into two metastable sub-levels with different population and
depopulation rates (zero-field splitting)
\cite{Basche-Nature-1995}. Each electronic level 
is characterized by a host of vibrational degrees of freedom
(cf. Fig.~1(a)). These vibrational states have a life time of about
$10^{-12}\,\mt{s}$ \cite{Basche-Nature-1995}. On the other hand, a transition from $\ket{3}$ to the
vibrational states of $\ket{1}$ occur on a time scale of $10^{-8}\,\mt{s}$ and
from $\ket{3}$ to the triplet state $\ket{2}$ about every
$10^{-3}\,\mt{s}$. Therefore these transitions can be described by effective
electronic Einstein coefficients $A_{31}, A_{32}^{(i)}, A_{21}^{(i)},
i = 1,2$, and by the Rabi frequency $\Omega_{31}$ of the laser driving
the $\ket{1} \leftrightarrow \ket{3}$ transition, while the vibrational levels
can be neglected. 
\begin{figure}
\includegraphics[width=8.5cm]{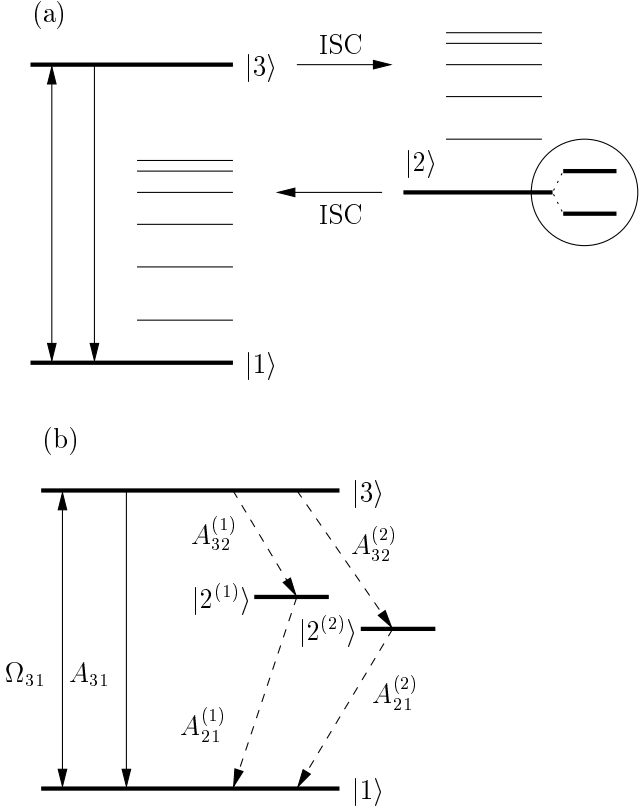}
\caption{\label{fig:terry-schema}Level scheme of terrylene. (a) Singlet
      ground state $\ket{1}$, excited singlet state $\ket{3}$, and
      triplet state $\ket{2}$, associated with a host
      of vibrational states. ISC (intersystem crossing,
      singlet-triplet transition) with low probability.
      Zero-field splitting of the 
      metastable triplet state into two sublevels with
      different population and depopulation rates. (b) The 
      extremely fast relaxation of the vibrational levels leads to
      an effective
      four-level system with two metastable states, $\ket{2^{(1)}}$
      and $\ket{2^{(2)}}$, and effective Einstein coefficients
      $A_{21}^{(i)},~ A_{32}^{(i)}~\ll~ A_{31},~ \Omega_{31}$. }
\end{figure}
Hence the level configuration effectively consists of a
four-level system as in Fig.~1(b), with two metastable states
$\ket{2^{(1)}}$ and $\ket{2^{(2)}}$, and one has 
\begin{equation} \label{eq:delta_t}
  A_{21}^{(i)}, A_{32}^{(i)} \ll A_{31}, \Omega_{31}~,\qquad i=1,2~.
\end{equation}
A fluorescence trajectory of such a system consists of light and dark
periods. During a light period, the subsystem $\{\ket{1}, \ket{3}\}$
behaves like a two-level system whose intensity is given by 
\cite{Carmichael-JP-1976} 
\begin{equation} \label{eq:Iss_2N}
  I_{\mt{L}} = \frac{A_{31}\Omega_{31}^2}{A_{31}^2 + 2\Omega_{31}^2}~.
\end{equation}
A transition from $\ket{3}$ to one of the metastable states $\ket{2^{(1)}}$ or
$\ket{2^{(2)}}$ each initiates a dark period (denoted by $\mt{D}^{(1)}$ and
$\mt{D}^{(2)}$) of mean 
duration $A_{21}^{(1)}$ and $A_{21}^{(2)}$, respectively. The
transition rates from the light period to the two dark periods are
denoted by $p_{\mt{LD}}^{(1)}$ and $p_{\mt{LD}}^{(2)}$, and those from the dark
periods to the light period by $p_{\mt{DL}}^{(1)}$ and
$p_{\mt{DL}}^{(2)}$. These transition rates can be calculated in a
fully quantum mechanical way by the
methods of Ref.~\cite{Addicks-EPJ-2001}, and they are given by (cf.~Appendix~A)
\begin{eqnarray} \label{eq:p_ij1}
p_{\mt{DL}}^{(i)} &=& A_{21}^{(i)}\\ \label{eq:p_ij2}
p_{\mt{LD}}^{(i)} &=& \frac{A_{32}^{(i)} \Omega_{31}^2}{A_{31}^2 +
  \Omega_{31}^2}~~. 
\end{eqnarray}
A fluorescence trajectory of terrylene is a Markovian jump process with
three intensity steps, i.e.~light, dark $\mt{D}^{(1)}$ and dark
$\mt{D}^{(2)}$. Recently, the present authors
\cite{Hegerfeldt-QSO-2002} have derived a highly accurate expression
for the intensity correlation function of arbitrary blinking systems
which is given in Appendix B and takes here the form
\begin{equation} \label{eq:gt}
  g(\tau) = \frac{1}{P_{\mt{L}}}\, P_{\mt{LL}}(\tau)\, g_2(\tau)~,
\end{equation}
where $P_{\mt{LL}}(\tau)$ is the conditional probability to have a light
period at time $t=\tau$ provided that at $t=0$ there was also a
light period, where 
\begin{equation} \label{eq:PL}
 P_{\mt{L}} = \lim_{\tau\rightarrow\infty} P_{\mt{LL}}(\tau) 
\end{equation}
is the probability to find a light period at all, and where $g_2(\tau)$
is the intensity correlation function of the $\{\ket{1}, \ket{3}\}$ two-level
subsystem. From Ref.~\cite{Carmichael-JP-1976} one has
\begin{equation} \label{eq:gt_2N}
  g_2(\tau) = 1 - \e^{\textstyle -\frac{3}{4}A_{31}\tau}\left(\cos\gamma\tau +
    \frac{3A_{31}}{4\gamma} \sin\gamma\tau\right)
\end{equation}
where
\begin{equation}
  \gamma=\frac{1}{4}\sqrt{16\Omega_{31}^2 - A_{31}^2}~.
\end{equation}
To determine $P_{\mt{LL}}(\tau)$ we consider the conditional probability,
$P_{\mt{LD}}^{(i)}(\tau)$, that at time $t=\tau$ one has a dark period
$\mt{D}^{(i)}$ under the condition of a light period at $t=0$ ($i=1,2$).
These conditional probabilities satisfy the rate equations
\begin{eqnarray}
\dot{P}_{\mt{LL}}(\tau) &=& \sum_{\alpha=1}^2 (-p_{\mt{LD}}^{(\alpha)}
P_{\mt{LL}}(\tau) + p_{\mt{DL}}^{(\alpha)}
P_{\mt{LD}}^{(\alpha)}(\tau)) \\
\dot{P}_{\mt{LD}}^{(i)}(\tau) &=& -p_{\mt{DL}}^{(i)}
P_{\mt{LD}}^{(i)}(\tau) + p_{\mt{LD}}^{(i)} P_{\mt{LL}}(\tau)~,~i=1,2
\end{eqnarray}
which is easily seen by noting that the first terms on the
r.h.s.~describe the decrease of the probabilities and the second
their increase. This can easily be generalized for an arbitrary number
of light and dark periods as shown in Appendix B. With the initial
condition $P_{\mt{LL}}(0)=1$ and $P_{\mt{LD}}^{(i)}(0)=0$ one finds in
a straightforward way 
\begin{eqnarray}
  P_{\mt{LL}}(\tau) &=&
  \frac{p_{\mt{DL}}^{(1)}p_{\mt{DL}}^{(2)}}{\mu_1\mu_2} -
\frac{\e^{\mu_1\tau}}{\mu_1(\mu_1-\mu_2)}\nonumber\\
&&\times\Bigl(p_{\mt{LD}}^{(1)}
(p_{\mt{DL}}^{(2)}+\mu_1) 
+ p_{\mt{LD}}^{(2)}(p_{\mt{DL}}^{(1)}+\mu_1)\Bigr) \nonumber\\ 
&&+ \frac{\e^{\mu_2\tau}}{\mu_2(\mu_1-\mu_2)}\nonumber\\
&&\times\Bigl(p_{\mt{LD}}^{(1)}(p_{\mt{DL}}^{(2)}+\mu_2) 
+ p_{\mt{LD}}^{(2)}(p_{\mt{DL}}^{(1)}+\mu_2)\Bigr)~.
\end{eqnarray}
where
\begin{eqnarray}
  \mu_{1,2} &=& -\frac{1}{2}(p_{\mt{DL}}^{(1)}+p_{\mt{LD}}^{(1)} +
  p_{\mt{LD}}^{(2)}+p_{\mt{DL}}^{(2)})\nonumber\\
  &&\pm   
\frac{1}{2}\left[(p_{\mt{DL}}^{(1)}+p_{\mt{LD}}^{(1)}-p_{\mt{LD}}^{(2)}-p_{\mt{DL}}^{(2)})^2
  +4p_{\mt{LD}}^{(1)}p_{\mt{LD}}^{(2)}\right]^{\frac{1}{2}}.\nonumber
\\
&&
\end{eqnarray}
From this and from Eq.~(\ref{eq:PL}) one obtains 
\begin{equation}
  P_{\mt{L}} = \frac{p_{\mt{DL}}^{(1)}p_{\mt{DL}}^{(2)}}{\mu_1\mu_2} =
  \frac{p_{\mt{DL}}^{(1)}p_{\mt{DL}}^{(2)}}{p_{\mt{LD}}^{(1)}p_{\mt{DL}}^{(2)} + p_{\mt{DL}}^{(1)}p_{\mt{LD}}^{(2)}+p_{\mt{DL}}^{(1)}p_{\mt{DL}}^{(2)}}~.
\end{equation}
Thus the correlation function $g(\tau)$ in Eq.~(\ref{eq:gt}) for
the four-level system of Fig.~1(b) satisfying the inequalities in
Eq.~(\ref{eq:delta_t}) is given in terms of the photo-physical
parameters $A_{31},\Omega_{31}, A_{32}^{(1)}, A_{32}^{(2)},
A_{21}^{(1)}$, and $A_{21}^{(2)}$.

One can re-express $g(\tau)$ through the mean durations of the three
periods, denoted by $T_{\mt{L}}, T_{\mt{D}}^{(1)}$ and
$T_{\mt{D}}^{(2)}$, respectively, and by the branching ratios
\begin{equation} \label{eq:pi}
  p_i = \frac{p_{\mt{LD}}^{(i)}}{p_{\mt{LD}}^{(1)} +
p_{\mt{LD}}^{(2)}}
\end{equation}
at the end of a light period, using \cite{Addicks-EPJ-2001}
\begin{eqnarray} \label{eq:TL}
  T_{\mt{L}} &=& \frac{1}{p_{\mt{LD}}^{(1)} +p_{\mt{LD}}^{(2)}}\\ \label{eq:TD}
  T_{\mt{D}}^{(i)} &=& \frac{1}{p_{\mt{DL}}^{(i)}},\quad i=1,2~.
\end{eqnarray}
With these relations one obtains, after some calculation, the explicit
expression
\begin{widetext}
\begin{eqnarray} \label{eq:gt_ausf}
g(\tau) &=&
g_2(\tau)~\left\{ 1+
  \frac{1}{T_{\mt{L}}}\exp\left[-\frac{1}{2}\left(\frac{1}{T_{\mt{L}}} +
    \frac{1}{T_{\mt{D}}^{(1)}} + \frac{1}{T_{\mt{D}}^{(2)}}\right)\tau\right]
  \Biggl[(p_1 T_{\mt{D}}^{(1)} + p_2 T_{\mt{D}}^{(2)})\cosh\Gamma\tau
\right.\nonumber\\
&&\left. + \Biggl(p_1 \left(\frac{T_{\mt{D}}^{(1)}}{T_{\mt{D}}^{(2)}}
  - \frac{T_{\mt{D}}^{(1)}}{T_{\mt{L}}} \right) + p_2\left(
       \frac{T_{\mt{D}}^{(2)}}{T_{\mt{D}}^{(1)}}-
       \frac{T_{\mt{D}}^{(2)}}{T_\mt{L}} \right) - 1 \Biggr)
      \frac{\sinh\Gamma\tau}{2\Gamma}\Biggr]\right\}~.   
\end{eqnarray}
\end{widetext}
where
\begin{equation}
  \Gamma = \frac{1}{2}\left((1/T_{\mt{D}}^{(1)} - 1/T_{\mt{D}}^{(2)} +
    (1-2p_2)/T_{\mt{L}})^2 + 4p_1 p_2/T_{\mt{L}}^2\right)^{\frac{1}{2}}.
\end{equation}
One sees that the statistics of intensity periods leads to the well known
bi-exponential tail in the intensity correlation function at a much
larger time scale than where $g_2(\tau)$ plays a role. Therefore one
has a hump at intermediate times $\tau\sim 10^{-7}\,\mt{s}$, where
$g_2(\tau)\simeq 1$ and $P_{\mt{LL}}(\tau)\simeq 1$. The height of
this hump is given by $1/P_{\mt{L}}$, according to Eq.~(\ref{eq:gt})
and to Eq.~(\ref{eq:gt_ausf}) for $\tau\to 0$ and
$g_2(\tau)\simeq 1$. Moreover, with the explicit form of $g(\tau)$ in
Eq.~(\ref{eq:gt_ausf}), all photo-physical parameters can be obtained
by a single fit to experimental data, as shown in the following.

\section{Example: Fit and results for terrylene}

We use the experimental data for the intensity correlation function of
terrylene obtained in Ref.~\cite{Basche-BBG-1996}. Details of the experimental setup can
be found in Ref.~\cite{Kummer-diss}. Since some of the laser light is
 scattered off 
the embedding crystal, with intensity $I_{\mt{sc}}$ , say,  this gives
rise to additional poissonian correlations. Therefore the intensity
correlation function in a light period is modified to the weighted
average
\begin{eqnarray}
  g_2^{\mt{mod}}(\tau) &=& \frac{I_{\mt{L}}g_2(\tau) +
    I_{\mt{sc}}}{I_{\mt{L}} + I_{\mt{sc}}}\\
  &=& 1 - \frac{\e^{\textstyle
      -\frac{3}{4}A_{31}\tau}}{1+I_{\mt{sc}}/I_{\mt{L}}}\left(\cos\gamma\tau +     \frac{3A_{31}}{4\gamma} \sin\gamma\tau\right)\nonumber\\
  &&
\end{eqnarray}
with $I_{\mt{L}}$ given by Eq.~(\ref{eq:Iss_2N}). This has to be inserted into
Eq.~(\ref{eq:gt_ausf}) instead of $g_2(\tau)$.

The thus modified Eq.~(\ref{eq:gt_ausf}) has been fitted to the
experimental data. In
Eq.~(\ref{eq:gt_ausf}) there are two different time scales -- for times $\tau >
10^{-7}\,\mt{s}$ the first factor, $g_2^{\mt{mod}}(\tau)$ is
essentially $1$ and for times $\tau < 10^{-7}\,\mt{s}$ the second factor
is practically constant. Therefore, we have first fitted the second
factor for $\tau > 10^{-7}\,\mt{s}$ and then fitted
$g_2^{\mt{mod}}(\tau)$ multiplied by a constant for $\tau <
10^{-7}\,\mt{s}$. Thus, in each of the two fits one has 
at most four free parameters, which reduces the numerical error. The
fitted curve and the experimental data are shown in Fig.~2, and the
resulting parameter values are listed in Table~I.
\begin{figure}
\includegraphics[width=8.5cm]{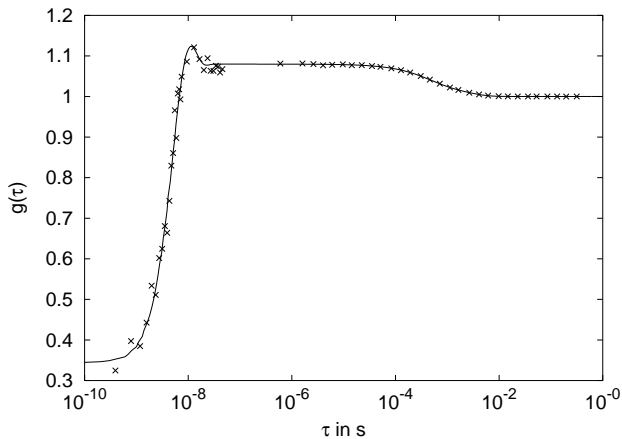}
\caption{\label{fig:anpassung}The intensity correlation function of a single
      fluorescent terrylene molecule embedded in a host crystal of
      p-terphenyl. Shown are the experimental data from 
      Ref.~\cite{Basche-BBG-1996} ($\times$) and the fit function
      obtained with Eq. (\ref{eq:gt_ausf}) (solid line). This function
      contains both the Rabi oscillations for small times as well as
      the long time correlations (hump) due to the metastable levels.}
\end{figure}
\setlength{\tabcolsep}{3.1mm}
\begin{table*}
\caption{\label{tab:parameter1}Photo-physical parameters for terrylene in
      p-terphenyl, obtained by a fit of Eq.~(\ref{eq:gt_ausf}) to the
      data of  Ref.~\cite{Basche-BBG-1996} and compared with results in
      the literature.}
\begin{ruledtabular}
\begin{tabular}{lccccccc}
 &$A_{31}$ & $\Omega_{31}$ & $I_{\mt{sc}}$  & $T_{\mt{L}}$& 
      $T_{\mt{D}}^{(1)}$& 
      $T_{\mt{D}}^{(2)}$& $p_1$  \\
      & $(10^8\,\mt{s}^{-1})$ & $(10^8\,\mt{s}^{-1})$ &
      $(10^7\,\mt{s}^{-1})$& (ms) & (ms) & (ms) & \\
      \hline 
      This work &
      $3.3\pm 0.5$ &
      $2.9\pm 0.1$ &
      $7.7\pm 0.7$ &
      $8.2\pm 0.3$ &
      $2.3\pm 0.2$ &
      $0.42\pm 0.03$ &
      $0.12 \pm 0.02$ \\ 
      Ref.~\cite{Basche-BBG-1996}&--&--&--&$6.3$&$3$&$0.43$&--\\ 
      Ref.~\cite{Kummer-diss}&2.9&--&--&$6.3\pm 0.1$&$3\pm 1$& 
      $0.43\pm0.01$ &--\\ 
      Ref.~\cite{Basche-Nature-1995}&--&--&--&$4.5$ &--&$0.4$&--     
\end{tabular}
\end{ruledtabular}
\end{table*}

 With the results of Table~I, one now could  calculate $A_{32}^{(i)}$
 and $A_{21}^{(i)}$ from 
Eqs.~(\ref{eq:p_ij1}), (\ref{eq:p_ij2}), and (\ref{eq:pi}) -
(\ref{eq:TD}). However, numerically it is
better to insert the expressions for $T_{\mt{L}}, T_{\mt{D}}^{(i)}$
and $p_i$ into Eq.~(\ref{eq:gt_ausf}) and then again fit the second factor
to the experimental data for $\tau > 10^{-7}\,\mt{s}$. With this
procedure one obtains all Einstein coefficients of the intersystem
crossings, as listed in Table~II.
\setlength{\tabcolsep}{2.6mm} 
\begin{table}
\caption{\label{}Einstein coefficients obtained here for the singlet-triplet
    transitions (ISC), compared with values in the literature.}
\begin{ruledtabular}
\begin{tabular}{lcccc}
 & $A_{21}^{(1)}$& $A_{21}^{(2)}$& $A_{32}^{(1)}$ & $A_{32}^{(2)}$ \\
      & $(\mt{s}^{-1})$& $(\mt{s}^{-1})$& $(\mt{s}^{-1})$& $(\mt{s}^{-1})$\\ 
      \hline 
      This work & 
      $430\pm 40$ &
      $2400\pm 200$ &
      $34\pm 7$ &
      $249\pm 8$ \\ 
      Ref.~\cite{Kummer-diss}&
      $250$ &
      $2300$ &
      $20$ &
      $290$ \\ 
      Ref.~\cite{Basche-BBG-1996}&
      $300$ &
      $2300$ &
      -- &
      $400$ \\ 
      Ref.~\cite{Basche-Nature-1995}&
      -- &
      $2500$ &
       -- &
      $400$
\end{tabular}
\end{ruledtabular}
\end{table}

The mean durations of the light and dark periods obtained here
essentially agree with those of Ref.~\cite{Basche-BBG-1996, Kummer-diss}, with a difference of
about $23\%$ in the case of $T_{\mt{L}}$ and with a drastically
reduced error bar in our 
value for $T_{\mt{D}}^{(1)}$. The reason probably is that a direct
determination of the period duration as done in Refs.~\cite{Basche-BBG-1996, Kummer-diss} depends
sensitively on the choice of the time window employed \cite{Addicks-EPJ-2001}. Our value
for $A_{31}$ agrees with that of Refs.~\cite{Basche-BBG-1996, Kummer-diss} within the error bars.

In case of the Einstein coefficients of the intersystem crossings our
values qualitatively agree with those of
Ref.~\cite{Basche-Nature-1995, Kummer-diss, Basche-BBG-1996}, where the
results of Refs.~\cite{Basche-BBG-1996, Kummer-diss} depend on the
same data as ours. There is 
overall agreement in so far as the population rates $A_{32}^{(i)}$ of
the metastable states are an order of magnitude smaller than the
corresponding depopulation rates $A_{21}^{(i)}$ and that the values
for $i=1$ are an order of magnitude smaller than those for $i=2$.
As a consequence, the metastable state $\ket{2^{(1)}}$ is occupied
considerably less often than $\ket{2^{(2)}}$, a fact also
mirrored in the largely different branching ratios $p_1$ and
$p_2$. Hence the dark period associated with the state $\ket{2^{(1)}}$
occurs much less frequently than the other one, and hence a statistical
determination may be more error prone than our fit procedure.
In Table II, somewhat larger differences are seen for $A_{21}^{(1)}$,
the population rate of the less frequently occupied metastable
state. 

\section{Conclusions and Outlook}

In this paper an explicit expression for the intensity correlation
function of blinking systems has been 
applied to a four-level system with two metastable states. Using
terrylene in p-terphenyl as an example it has been shown that the
relevant photo-physical parameters can be obtained from a single fit to
experimental data for the correlation function. This is a considerable
simplification compared to previous procedures, and it can also result
in smaller error bars. In this way we have obtained the Einstein
coefficient of the fast transition, the mean durations of the light
and dark periods, as well as the population and depopulation rates of
the metastable levels. Our results qualitatively agree with values
given in the literature. In addition we have obtained further
parameters like the Rabi frequency $\Omega_{31}$ and  the
contribution, $I_{\mt{sc}}$, to the correlation function due to the light
back-scattered off the crystal.

It should also be possible to apply our approach of an explicit
calculation of the correlation function not only to terrylene but also
to other blinking systems with a more complicated statistics of light and dark periods. Particularly interesting in this context would be the 
recently performed fluorescence measurement of two dipole-interacting 
molecules \cite{Hettich} and to determine the interaction parameters
of such a system by the present methods.

\appendix
\section{}

We indicate here the derivation of Eqs.~(\ref{eq:p_ij1}) and
(\ref{eq:p_ij2}) for the 
transition rates between the different fluorescence periods of the
level system in Fig.~1(b). To do this we use the method of
Ref.~\cite{Addicks-EPJ-2001}. During a dark period $\mt{D}^{(i)}$, the
system is, to high accuracy, in the 
state $\ket{2^{(i)}}$, with corresponding density matrix
\begin{equation}
  \rho_{\mt{D}}^{(i)} = \ket{2^{(i)}}\bra{2^{(i)}}~,
\end{equation}
and during a light period in the steady state of the $\{\ket{1},\ket{3}\}$
subsystem, with the density matrix
\begin{eqnarray}
  \rho_{\mt{L}} &=& \frac{1}{A_{31}^2 + 2\Omega_{31}^2}\Bigl(
  (A_{31}^2 + \Omega_{31}^2)\ket{1}\bra{1}\nonumber\\
  &&+ \Omega_{31}^2
  \ket{3}\bra{3} + \i A_{31}\Omega_{31}(\ket{1}\bra{3} -
  \ket{3}\bra{1})\Bigr)~.
\end{eqnarray}
Starting from one of these density matrices, the increase in
occupation of the other subsystems during a time $\Delta t$ yields the
transition rates; here ${\Delta t}$ has to satisfy
\begin{equation} \label{eq:delta_t_app}
  (A_{32}^{(i)})^{-1}, (A_{21}^{(i)})^{-1} \ll \Delta t \ll
  A_{31}^{-1}, \Omega_{31}^{-1}~.
\end{equation}
With initial state $\rho(t_0) = \rho_{\mt{L}}$ one then has
\begin{equation} \label{eq:pLD}
  p_{\mt{LD}}^{(i)} = \frac{\d}{\d
    t}\bra{2^{(i)}}\rho(t)\ket{2^{(i)}}_{t=t_0+\Delta t}
\end{equation}
and for $\rho(t_0)=\rho_{\mt{D}}^{(i)}$ one has
\begin{equation} \label{eq:pDL}
  p_{\mt{DL}}^{(i)} = \frac{\d}{\d t} \Bigl(\bra{3}\rho\ket{3} +
\bra{1} \rho \ket{1} \Bigr)_{t=t_0+\Delta t}~.
\end{equation}
To evaluate this we use the Bloch equations which can be written in
the form \cite{Hegerfeldt-PRA-1993, Hegerfeldt-QSO-1996}
\begin{eqnarray} \label{eq:blochgleichung}
  \frac{\d}{\d t}\rho(t) &=& -\frac{\i}{\hbar}
  \left(H_{\mt{C}}\rho(t)-\rho(t)H_{\mt{C}}^{\dagger}\right) +
  \cal{R}(\rho(t))\\
  &\equiv& \L \rho(t)~ \label{eq:bloch_L},
\end{eqnarray}
where the so-called conditional Hamiltonians $H_{\mt{C}}$ is given by
\begin{eqnarray}
  H_{\mt{C}} &=&  \frac{\hbar}{2\i}\Bigl(A\ket{3}\bra{3}
  + A_{21}^{(1)}\ket{2^{(1)}}\bra{2^{(1)}} +
  A_{21}^{(2)}\ket{2^{(2)}}\bra{2^{(2)}}\Bigr)\nonumber\\
  &&+
  \frac{\hbar\Omega_{31}}{2}\Bigl(\ket{1}\bra{3} + \ket{3}\bra{1}\Bigr)
\end{eqnarray}
and the super-operator $\cal{R}$ by
\begin{eqnarray} \label{eq:ruecksetz}
  \cal{R}(\rho) &=& \sum_{i=1}^2 \ket{1} \Bigl(
  A_{31}\bra{3}\rho\ket{3} + A_{21}^{(i)}\bra{2^{(i)}} \rho
  \ket{2^{(i)}} \Bigr)\bra{1}\nonumber\\
&&+ A_{32}^{(i)}\ket{2^{(i)}}\bra{3} \rho \ket{3} \bra{2^{(i)}}~.
\end{eqnarray}
Using Eq.~(\ref{eq:blochgleichung}) one can show that Eqs.~(\ref{eq:pLD})
and (\ref{eq:pDL}) can be written as
\begin{eqnarray}
 p_{\mt{LD}}^{(i)} &=& A_{32}^{(i)}\bra{3}\rho(t_0 + \Delta t)\ket{3} -
 A_{21}^{(i)}\bra{2^{(i)}}\rho(t_0 + \Delta t)\ket{2^{(i)}}
 \nonumber\\&&\label{eq:pLD_rho} \\
 p_{\mt{DL}}^{(i)} &=& \sum_{\alpha=1}^2 \Bigl( A_{21}^{(\alpha)}
 \bra{2^{(\alpha)}} \rho(t_0 + \Delta t) \ket{2^{(\alpha)}}\nonumber\\
 &&-
 A_{32}^{(\alpha)} \bra{3} \rho(t_0 + \Delta t) \ket{3}
 \Bigr)~.\label{eq:pDL_rho}  
\end{eqnarray}
The necessary matrix elements of $\rho(t_0+\Delta t)$ are obtained
from the Bloch equations written as in of Eq.~(\ref{eq:bloch_L}). Writing
$\L$ in the obvious form
\begin{equation}
  \L = \L_0(A_{31},\Omega_{31}) + \L_1(A_{32}^{(i)}, A_{21}^{(i)})
\end{equation}
one obtains in a way analogous to the usual quantum mechanical
perturbation theory, to first order in $A_{32}^{(i)}$ and $A_{21}^{(i)}$, 
\begin{eqnarray}
  \rho(t_0+\Delta t) &=& \e^{\L \Delta t}\rho(t_0)\nonumber\\
  &=& e^{\L_0 \Delta t}\rho(t_0) + \int_0^{\Delta t}\!\!\!\!\!\! \d \tau~
  \e^{\L_0(\Delta t -\tau)} \L_1 \e^{\L_0\tau}\rho(t_0)~.\nonumber\\&&
\end{eqnarray}
Since $\rho (t_0)$ is a stationary state of subsystems one has
$\L_0\rho(t_0) = 0$ and hence $\e^{\L_0\tau}\rho(t_0)=\rho(t_0)$.
After a change of variable one thus obtains
\begin{equation} \label{eq:rho(t)}
 \rho(t_0+\Delta t) = \rho(t_0) + \int_0^{\Delta t} \d \tau~
 \e^{\L_0\tau} \L_1 \rho(t_0)~.
\end{equation}
Now $\L_1\rho(t_0)$ can be decomposed into a component parallel to the
null space of $\L_0$ and a component belonging to the subspace for
nonzero eigenvalues,
\begin{equation} \label{eq:zerleg_L1}
 \L_1\rho(t_0) = \proj_{\parallel} \L_1\rho(t_0) + (\eins- \proj_{\parallel})
\L_1\rho(t_0)~. 
\end{equation}
Here
\begin{equation} 
  \proj_{\parallel} = \frac{1}{2\pi \i} \oint_{\cal{C}_0}\d z\,(z-\L_0)^{-1}~,
\end{equation}
where $\cal{C}_0$ is a path in the complex plane enclosing no other
eigenvalue than zero \cite{Kato-book}. The integral can be
calculated by the residue method.
In view of Eq.~(\ref{eq:delta_t_app}) the contribution to
Eq.~(\ref{eq:rho(t)}) of the $\proj_{\parallel}$ term in 
Eq.~(\ref{eq:zerleg_L1}) is negligible. The contribution to
Eq.~(\ref{eq:rho(t)}) of the $(\eins- \proj_{\parallel})$ 
term   is governed by the
nonzero eigenvalues of $\L_0$ which possess a (large) negative
real part of the order of $A_{31}$ and $\Omega_{31}$. Therefore the
integrand in Eq.~(\ref{eq:rho(t)}) is strongly damped in $\tau$, and in view of
Eq.~(\ref{eq:delta_t_app}) the upper integration limit can be replaced by
$\infty$, yielding
\begin{eqnarray}
  \rho(t_0+\Delta t) &=& \rho(t_0) + \int_0^{\infty} \d \tau~
 \e^{\L_0\tau} (\eins-\proj_{\parallel})\L_1 \rho(t_0)\nonumber\\
 &=& \rho(t_0) + (\epsilon - \L_0)^{-1} (\eins - \proj_{\parallel}) \L_1 \rho(t_0)~,\nonumber\\&&
\end{eqnarray}
with $\epsilon \rightarrow +0$. Inserting this into
Eqs.~(\ref{eq:pDL_rho}) and (\ref{eq:pLD_rho}) one
obtains, after some calculations involving 16$\times$16 matrices, the
transition rates of Eqs.~(\ref{eq:p_ij1}) and (\ref{eq:p_ij2}). These hold
to first order in $A_{21}^{(i)}$ and $A_{32}^{(i)}$.

\section{}

We give here generalized expressions for the intensity correlation
function of fluorescing systems with an arbitrary number of different
intensity periods with intensity $I_i$~. A detailed discussion can be found in
Ref.~\cite{Hegerfeldt-QSO-2002}.

Let $P_i$ be the probability for the occurrence of period $i$
 and let $P_{ij}(\tau)$ be the probability to have period 
$j$ at time $\tau$ provided one had period $i$ at $\tau = 0$. Then it
can be shown, that a highly accurate expression for the normalized
intensity correlation function is 
\begin{equation} \label{eq:gt_ausf_app}
  g(\tau) = \frac{\sum_{ij} P_i
    I_i I_jP_{ij}(\tau) g_j(\tau)}{\left(\sum_{\alpha}
      P_{\alpha}I_{\alpha}\right)^2}~, 
\end{equation}
where $g_i(\tau)$ is the correlation function within a given period
$i$ and usually easier to calculate than that of the complete
system. In the case of only one light period and two dark periods
Eq.~(\ref{eq:gt_ausf_app}) reduces to Eq.~(\ref{eq:gt}).  

Let $p_{ij}$ be the transition rates from period $i$ to $j$. Then the $P_{ij}(\tau)$ are easily seen to obey rate equations, e.g.
\begin{equation}
  \dot{P}_{11}(\tau) = \Bigl(-\sum_k p_{1k}\Bigr)P_{11}(\tau) +
    p_{21}P_{12}(\tau)+ \cdots + p_{n1}P_{1n}(\tau)~.
\end{equation}
In general, with the matrix ${\bf B} = (B_{ij})$,
\begin{equation}
 B_{ij}=p_{ij} -
 \delta_{ij}\sum_{k}p_{ik}~,  
\end{equation}
and the matrix
\begin{equation}
  {\bf P}(\tau) = \Bigl(P_{ij}(\tau)\Bigr)
\end{equation}
one has
\begin{equation} \label{eq:P_diffeq}
 \dot{\bf P} = {\bf P}{\bf B}~, 
\end{equation}
with the initial condition $P_{ij}(0) = \delta_{ij}$, or
\begin{equation}
 {\bf P}(0) = \eins~. 
\end{equation}
The solution of equation (\ref{eq:P_diffeq}) with this initial condition
can be written as
\begin{equation} \label{eq:P_tau}
 {\bf P}(\tau) = \e^{{\bf B}\tau}~. 
\end{equation}
If $\mu_0, \dots, \mu_{n-1}$ are the eigenvalues of ${\bf B}$ (assumed
distinct) then \cite{Gant-book}
\begin{equation} \label{eq:gantm_formel}
  \e^{{\bf B}\tau} = \sum_{i=0}^{n-1}\e^{\mu_i\tau}\prod_{\alpha\neq
    i}\frac{{\bf B}-\mu_{\alpha}}{\mu_i-\mu_{\alpha}}~.
\end{equation}
The properties of the matrix ${\bf B}$ are closely related to those of
stochastic matrices \cite{Gant-book}, and under quite
general conditions ${\bf B}$ has a single eigenvalue $\mu_0 = 0$ and
eigenvalues $\mu_1,\dots, \mu_{n-1}$ with negative real part.

To find the $P_i$'s, we note that for $\tau \rightarrow \infty$ the
memory to the initial condition is in general lost. Therefore, for any
$\kappa$, 
\begin{equation} \label{eq:P_i}
  P_i = P_{\kappa i}(\infty)~.
\end{equation}

\bibliographystyle{apsrev}
\bibliography{705317JCP}

\end{document}